\documentclass[preprint,nocopyrightspace,numbers]{sigplanconf}

\usepackage{amsmath}
\usepackage{graphicx}

\begin{document}

\setlength{\pdfpageheight}{\paperheight}
\setlength{\pdfpagewidth}{\paperwidth}

\conferenceinfo{CONF 'yy}{Month d--d, 20yy, City, ST, Country}
\copyrightyear{20yy}
\copyrightdata{978-1-nnnn-nnnn-n/yy/mm}
\copyrightdoi{nnnnnnn.nnnnnnn}


\preprintfooter{PROHA'16, March 12, 2016, Barcelona, Spain}

\title{TANGO: Transparent heterogeneous hardware Architecture deployment for eNergy Gain in Operation}

\authorinfo{K. Djemame \and D. Armstrong \and R. Kavanagh}
           {University of Leeds, UK}
           {K.Djemame@leeds.ac.uk}
\authorinfo{J.C. Deprez}
           {CETIC, Belgium}
           {Jean-Christophe.Deprez@cetic.be}
\authorinfo{A.J. Ferrer \and D. Garcia Perez}
{Atos, Spain}
{ana.juanf@atos.net}
\authorinfo{R.M. Badia \and R. Sirvent \and J. Ejarque}
{Barcelona Supercomputing Center, Spain}
{R.M.Badia@bsc.es}
\authorinfo{Y. Georgiou}
{Bull, France}
{yiannis.georgiou@atos.net}

\maketitle

\begin{abstract}
The paper is concerned with the issue of how software systems actually use Heterogeneous Parallel Architectures (HPAs), with the goal of optimizing power consumption on these resources. It argues the need for novel methods and tools to support software developers aiming to optimise power consumption resulting from designing, developing, deploying and running software on HPAs, while maintaining other quality aspects of software to adequate and agreed levels. To do so, a reference architecture to support energy efficiency at application construction, deployment, and operation is discussed, as well as its implementation and evaluation plans.
\end{abstract}

\category{C2.4}{Distributed Systems}{Distributed Applications} 



\section{Introduction}

Recent years saw the emergence of Cyber-Physical Systems (CPS), the Internet of Things (IoT), and the Smart Anything Everywhere Initiative which have the potential to transform the way we live and work \cite{EC:2015}. For example, the IoT transformational impact in the long term is expected to increase significantly with mass adoption, tens of billions of things connected, and a multi-trillion-dollar economic value, the key drivers being new business models taking advantage of the data collected by the IoT, sophisticated application development platforms, analytics applied to things, and distributed/parallel architectures \cite{Gartner:2015}.

As the range of applications continues to grow, e.g. CPS, IoT, connected smart objects, High Performance Computing (HPC), mobile computing, wearable computing etc. there is an urgent need to design more flexible software abstractions and improved system architectures to fully exploit the benefits of the heterogeneous platforms on which they operate. Heterogeneous parallel architectures have received considerable attention, as an efficient approach to run applications and deliver services, by combining different processor types in one system to improve absolute performance, minimise power consumption and/or lower cost. New platforms incorporating multi-core CPUs, many-core GPUs, and a range of additional devices into a single solution are being introduced. These platforms are showing up in a wide range of environments spanning supercomputers to personal smartphones.  One of the challenges to future application performance lies with not only efficient node-level execution but power consumption as well, which is a key focal point of this paper.

Although general complex engineering simulations come to mind when identifying families of applications benefiting most from heterogeneous parallel architectures, in the upcoming era of IoT and Big Data, there is significant interest in exploiting the capabilities offered by customised heterogeneous hardware such as FPGA, ASIP, MPSoC, heterogeneous CPU+GPU chips and heterogeneous multi-processor clusters all of which with various memory hierarchies, size and access performance properties. In fact, Big Data online with nearly instantaneous results demand massive parallelism and well devised divide-and-conquer approaches to exploit heterogeneous hardware, both client and server sides, to its fullest extent. Moreover, heterogeneous systems can not only handle workload with fewer and/or smaller servers (cost saving) but also slash the energy used to run certain applications, which helps gain clear benefits and addresses the growing interest in green solutions and the pressure to reduce the environmental impact of, e.g. data centres. A common theme across all scenarios is the need for low-power computing systems that are fully interconnected, self-aware, context-aware and self-optimising within application boundaries \cite{Bortolotti:2013}. 

Because the impact of heterogeneity on all computing tasks is rapidly increasing, innovative architectures, algorithms, and specialized programming environments and tools are needed to efficiently use these new and mixed/diversified parallel architectures. 

The paper is concerned with low-power multi / many-core/ programmable computing systems development by addressing the power consumption and efficiency of the software which runs on these infrastructures. As software consumes energy in its operation, the primary aim of this research is to relate software design and power consumption awareness, making it imperative that the software to be developed is not only as low power consumption aware as it possibly can be, but takes into account trade-offs with other key requirements in the environment where it runs such as performance, time-criticality, dependability, data movement, security and cost-effectiveness as well.

The paper's main contributions are: 1) the incorporation of a novel approach that combines energy-awareness related to heterogeneous parallel architectures with the principles of requirements engineering and design modelling for self-adaptive software-intensive systems. This way, the energy efficiency of both heterogeneous infrastructures and software are considered in the application development and operation lifecycle, and 2) a proposed energy efficiency aware system architecture, its components, and their roles to support key requirements in the environment where it runs such as performance, time-criticality, dependability, data movement, security and cost-effectiveness. 

The remainder of the paper is structured as follows: Section II describes the proposed architecture to support energy-awareness. Section III discusses optimisation issues supported through self-adaptation to enact optimal, in terms of requirements and Key Performance Indicators (KPIs), application deployment and operation. Section IV discusses the implementation and evaluation plans of the architecture. Section V presents related work. In conclusion, Section VI provides a summary of the research and plans for future work.

\begin{figure*}[ht]
	\centering
	\includegraphics[width=0.8\textwidth]{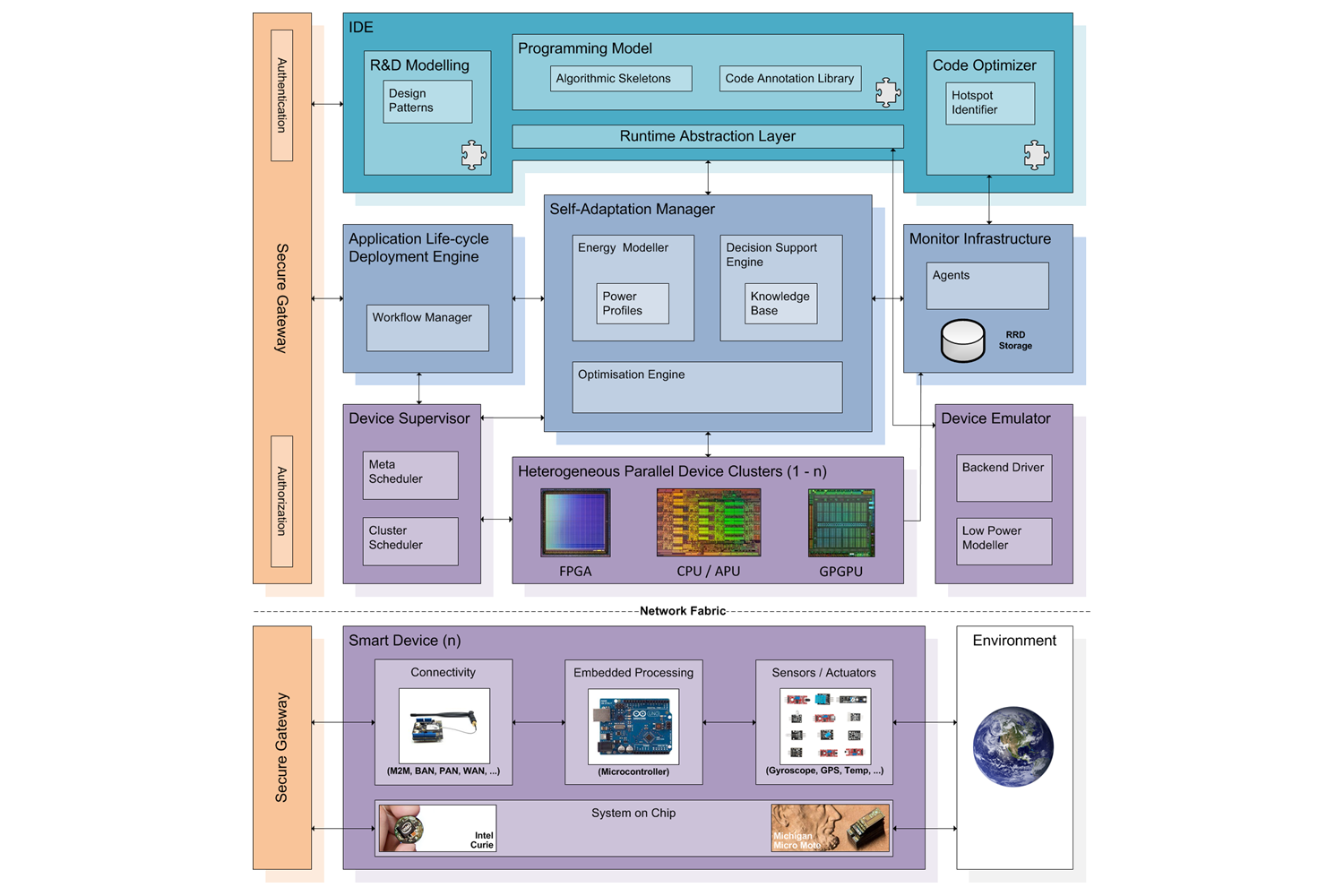}
	\caption{Reference Architecture}
	\label{fig:arch}
\end{figure*}

\section{System Architecture}

As stated in Section 1, it is clear the energy requirements of the software applications which run on heterogeneous parallel architectures must be incorporated into the overall development and deployment process. Determining the relationship between software structure and its power usage will allow the definition of a set of software power metrics similar in concept to those for hardware. By associating those metrics with software components and libraries it will be possible to not only populate a software development environment with information to predict and illustrate the power requirements of applications enabling the programmer to see the consequences of their work, but also automatically optimise the code by allowing alternative selections of software components to be made \cite{Farokhi:2015}, using power consumption as an additional selection criterion. In addition to this, developers need to fully understand the nuances of different hardware configurations and software systems (both rapidly evolving), as well as consider additional issues in terms of performance, security mixed-criticality and power consumption resulting from the heterogeneous system.

Next, the proposed architecture in Figure \ref{fig:arch} is discussed in the context of the application life cycle: construction, deployment, and operation. It is separated into remote processing capabilities in the upper layers, which in turn is separated into distinct blocks that support the standard application deployment model (construct, deploy, run, monitor, adapt) and local processing capabilities in the lowest layer, which illustrates support for secure embedded management of IoT devices and associated I/O.

\subsection{Integrated Development Environment (IDE)}
In this block, a collection of components interact to facilitate the modelling, design and construction of applications. The components aid in evaluating power consumption of an application during its construction. A number of plug-ins are provided for a frontend IDE as a means for developers to interact with components within this layer. Lastly, this layer enables architecture agnostic deployment of the constructed application, while also maintaining low power consumption awareness. The components in this block are:

{\it Requirements and Design Modelling}: aims at guiding the devel-opment and configuration of applications to determine what can be targeted in terms of Quality of Service (QoS), Quality of Protection (QoP), cost of operation and power consumption behaviour when exploiting the potential of the underlying heterogeneous hardware devices. In particular, it is anticipated that different deployment alternatives of an application will lead to different levels of quality, power consumption behaviour and operational cost. The Requirements and Design Modelling tools must therefore help to better understand deployment alternatives in particular situations. Rapid prototyping with exploratory runs of an application or portion of an application on actual heterogeneous hardware devices or on device emulators will help developers with design decisions such as identify what software components to decouple from others so their execution can exploit the parallelism of the provided underlying heterogeneous hardware. It is also at this state that developers can determine what portion of an application could be compiled or rapidly ported to programmable hardware to further optimise trade-off on quality, energy and cost performance.

{\it Programming model (PM)}: supports developers when coding their applications.  Although complex applications are often written in a sequential fashion without clearly identified APIs, the PM let programmers annotate their programs in such a way that the Programming Model Runtime can then execute them in parallel on heterogeneous parallel architectures. At runtime, applications described for execution with the Programming Model runtime are aware of the power consumption of components implementation. In the task-based paradigm supported by the programming model, tasks are annotated by the developer, indicating directionality of the task parameters, and at runtime a task dependence graph is built which inherently describes the parallelism of the application. Heterogeneity is easily handled with this paradigm, since tasks that better fit a given device will be executed there, and locality aspects can also be easily taken into account by the runtime. A hierarchy of task-based programming models will be used in a project, combining coarser grain and finer grain tasks, which will enable on one hand to better capture the high level structure of the application (coarse grain) and the other hand, details of the architecture (finer grain). Different instances of tasks would be available, in such a way that at runtime the best one (in terms of energy or time selected trade-off) can be chosen and executed in the optimum device.

{\it Code optimizer}: plays an essential role in the reduction of energy consumed by an application. This is achieved through the adaptation of the software development process and by providing software developers the ability to directly understand the energy foot print of the code they write. The proposed novelty of this component is in its generic code based static analysis and energy profiling capabilities (Java, C, C++, etc. available in the discipline of mobile computing) that enables the energy assessment of code out-of-band of an application's normal operation within a developer's IDE.

\subsection{Application Deployment}
This block consists of a set of components to handle the placement of an application considering energy models on target heterogeneous parallel architectures. It aggregates the tools that are able to assess and predict performance and energy consumption of an application. Application level monitoring is also accommodated, in addition to support of self-adaptation for the purpose of making decisions using application level objectives given the current state of the application in question. The components in this block are:

{\it Application Life cycle Deployment Engine}: this component manages the lifecycle of an application deployed by the IDE. Once a deployment request is received, this component must choose the infrastructure that is most suitable according to various criteria, which include for example: 1) energy constraints/goals that indicate the minimum energy efficiency that is required/desired for the deployment and operation of an application; 2) application performance constraints that indicate the minimum requirements in terms of performance for the application (time-criticality, data location, cost etc.) This will be made possible through the enhanced heterogeneous resources description as implemented within the resource and job management system used, e.g. SLURM \cite{SLURM:2003}. The different application needs and criteria will be selected through the interface provided by SLURM. The enhanced SLURM will perform automatic workload execution upon the heterogeneous platform, in addition to managing data (stage-in, stage-out), by applying efficient scheduling techniques between jobs (fair sharing, backfilling, pre-emption, etc.) and by selecting the best-suited resources for each job (based on resources characteristics, network topology, internal node topology, power management, etc.). Moreover, this component's role is also to optimize the life cycle of an application to ensure its constraints are fulfilled considering: 1) the status of the heterogeneous parallel devices in terms of power consumption and workload; 2) the description of the cluster in terms of platform type, hardware specification and its power consumption profile, and 3) profile of application in terms of how it stresses each of the devices (CPU, memory, network ...). Using SLURM's support for heterogeneous resources, the accounting and profiling of each heterogeneous resource will take place for all jobs.

{\it Monitor Infrastructure}: this component is able to monitor the heterogeneous parallel devices (CPU, memory, network ...) that are being consumed by a given application by providing historical statistics for device metrics. The monitoring of an application must be performed in terms of power/energy consumed (e.g. Watts that an application requires during a given period of its execution), and performance (e.g. CPU that an application is consuming during a given period of its execution).

{\it Self-Adaptation Manager}: This component provides key functionality to manage the entire adaptation strategy applied to applications and Heterogeneous Parallel Devices (HPDs). This entails the dynamic optimisation of: energy efficiency, time-criticality, data movement and cost-effectiveness through continuous feedback to other components within the architecture and a set of architecture specific actuators that enable environmental change. Examples of such actuators could be: redeployment to another HPD, restructuring a workflow task graph or dynamic recompilation. Furthermore, the component provides functionality to guide the deployment of an application to a specific HPD through predictive energy modelling capabilities and polices, defined within a decision support engine, which specify cost constraints via Business Level Objectives (BLOs). More details are available in Section 3.

\begin{figure*}[ht]
	\centering
	\includegraphics[width=0.75\textwidth]{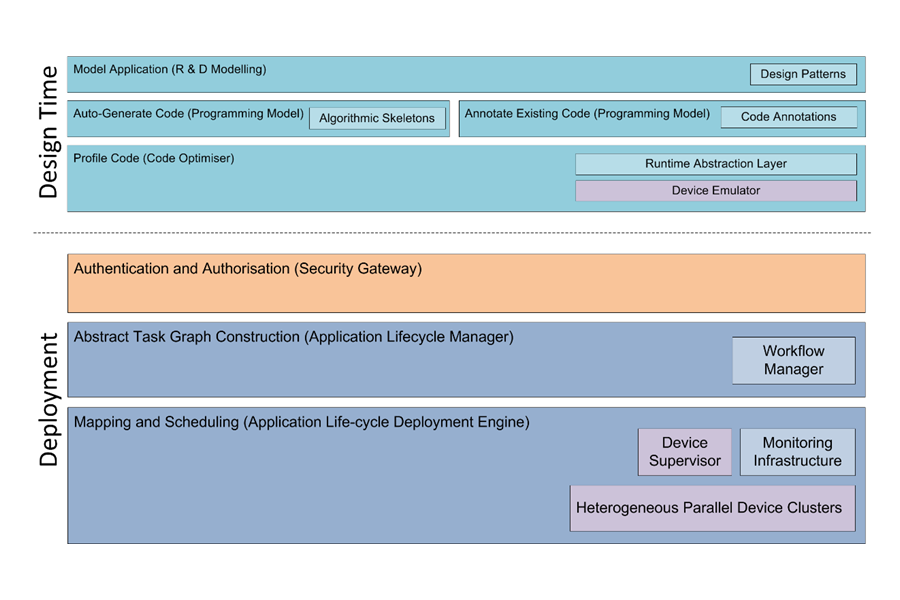}
	\caption{Architecture support for training application power profiles and deployment}
	\label{fig:arch-support}
\end{figure*}

\subsection{Heterogeneous Parallel Devices Management}
The last block above the network fabric line, addresses the heterogeneous parallel devices and their management. The application admission, allocation and management of HPDs are performed through the orchestration of a number of components. Power consumption is monitored, estimated and optimized using translated application level metrics. These metrics are gathered via a monitoring infrastructure and a number of software probes. At runtime HPDs will be continually monitored to give continuous feedback to the Self-Adaptation Manager. This will ensure the architecture adapts to changes in the current environment and in the demand for energy. Optimizations take into account several approaches, e.g. redeployment to another HPD, dynamic power management policies considering heterogeneous execution platforms and application energy models. The components in this block are:

{\it Device Supervisor}: This component provides scheduling capabilities across devices during application deployment and operation. This covers the scheduling of workloads of both clusters (Macro level, including distributed network and data management) and HPDs (Micro level, including memory hierarchy management). The component essentially realises abstract workload graphs, provided to it by the Application Life-cycle Deployment Engine component, by mapping tasks to appropriate HPDs. Meta-scheduling heuristics manage multiple clusters efficiently, while cluster level heuristics optimise the use of HPD resources and resource sets. Optimisation criteria (such as power consumption) and environment state are provided as input by the Self-Adaptation Manager and Monitoring Infrastructure components respectively.

{\it Device Emulator}: This component provides out-of-band application deployment and operation to emulated HPD resources for the purpose of training application power profiles. Emulated HPD resources execute application code while KPIs are monitored. The output of this process calibrates metrics within a power model that is provided to the Self-Adaptation Manager as a power profile, the normalised performance results of as running an application on a specific type or combination of HPD. Emulation of a range of HPDs is realised through a generic back end driver that interfaces to hardware emulators such as QEMU, OpenCL Emulator (ocl-emu) or vendor specific ASIC (FPGA) emulators. The device emulator could also be re-purposed to provide development time debugging capabilities.

Furthermore, a Secure Gateway supports pervasive authentication and authorization, which at the core of the proposed architecture enables both mobility and dynamic security. This protects components and thus applications from unauthorised access, which in turn improves the dependability of the architecture as a whole. The component provides embedded smart devices from the IoT paradigm, secure access to remote processing resources through the network fabric as well as enabling secure management of these devices through the upper layers of the architecture. These smart devices, comprised of a combination of discrete embedded components (providing connectivity, embedded processing, sensors, etc.) or a System on Chip (SoC), sense or actuate on an environment and filter acquired data using limited processing capabilities. After this local processing, data is sent securely over the network fabric for further remote processing on more capable heterogeneous parallel devices, supported by the upper layers of the architecture.

\begin{figure*}[ht]
	\centering
	\includegraphics[width=0.75\textwidth]{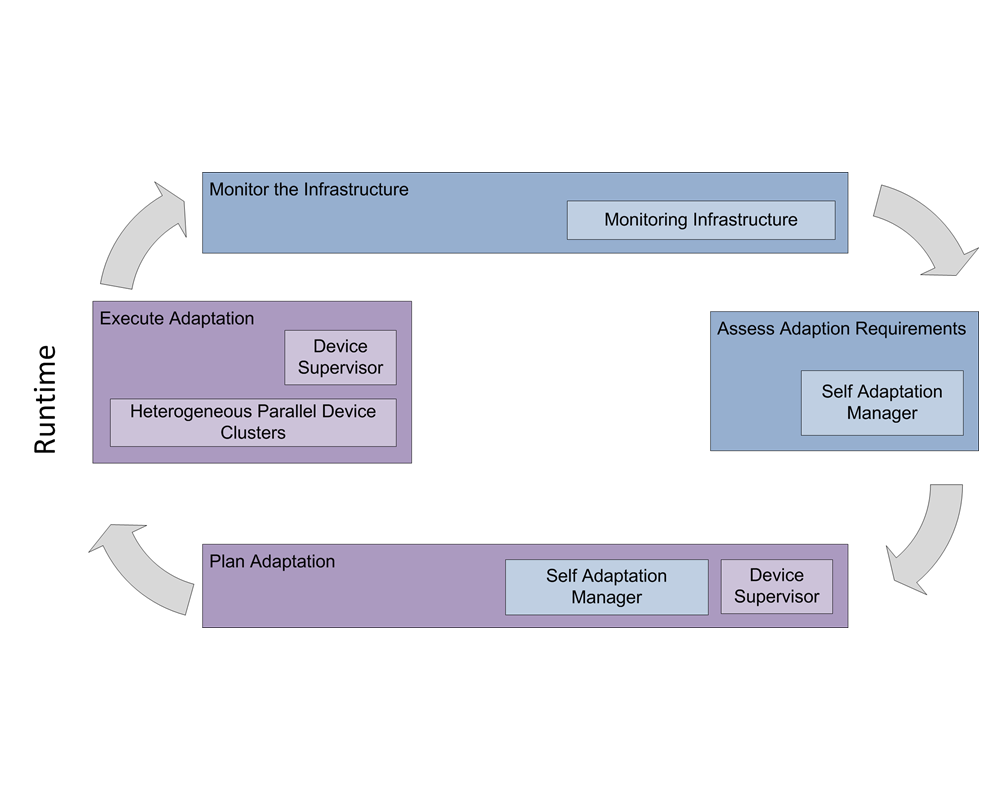}
	\caption{Architecture support for self-adaptation at runtime}
	\label{fig:arch-self-adapt}
\end{figure*}

\section{Self-Adaptation Support}
This section discusses the interactions of these components in the context power consumption and the ramifications this has on application design, deployment and adaptation.

The proposed architecture is designed to support: 1) application profiling at design time, and 2) actual deployment on heterogeneous parallel devices, as shown in Figure \ref{fig:arch-support}. This way, an application can be trained for a specific parallel architecture. In doing so, the Energy Modeller is able to create a better power consumption prediction once the application is deployed. Furthermore, programmers can identify portion of code worth re-writing and re-annotating to augment self-adaptability at runtime, e.g. to better exploit newly added hardware capabilities by the underlying programming model. 

In addition to support for application design and deployment, the proposed architecture provides capabilities to perform continuous autonomic self-adaptation during runtime as shown in Figure \ref{fig:arch-self-adapt}. This leverages fine-grained monitored metrics of heterogeneous parallel devices and application software to create an adaptation plan supporting the performance and cost goals of an application. It is achieved through advances in modelling and prototyping that enable power, cost and performance awareness during operation through emulation and simulation under various "what-if" scenarios.

\section{Implementation Plan}
This research will create a new cross-layer programming flow for heterogeneous parallel architectures featuring automatic code mapping and power optimization on the target architecture. Therefore, as explained in section 2, it will consider the following aspects: 1) application programming; 2) middleware, with support for power consumption modelling and power-aware emulation to support the actual code mapping, and 3) architecture(s) where the application runs.

Mapping complex software on heterogeneous multi-core hard-ware is performed as follows. The resource usage of the software will be modelled as software metadata, which can be data flow, Petri-net-based \cite{Jensen:2009} or any other relevant formalism. This software model is mapped to a similar hardware metadata representing the capabilities of the targeted hardware. Once this theoretical mapping is found, the software configuration is adapted to the targeted hardware, following the resulting mapping. This can involve adding data movement, notably to manage scratchpad memory, or selecting the proper compiler or parallelisation \cite{Pena:2014}. Software metadata can include a number of parameters ranging from the process description (fork, synchronization), memory requirements (allocation, free, communication), deadlines, switches annotated with their associated probabilities, etc. Besides, hardware metadata include a description of the hardware, together with its particular cost function (e.g. cost of data movement). Defining a model for software and hardware metadata may encompass a probabilistic annotation. Such annotation could be either expected pre-runtime, or could be inferred statistically by observing the behaviour of the software in operation in order to improve future optimisation decisions. The actual mapping is performed by means of optimization algorithms that optimize utility functions such as power consumption, end to end timing, cost etc. To this end, this research will use a generic optimization engine that is designed to support a realistic model during the optimization phase, OscaR \cite{OSCAR}. This will ensure that the delivered mapping is as close as possible to the actual optimum that will be encountered during the execution of the mapped software on the targeted hardware. If the solution is found to be sub-optimal after deployment, future optimisation will be enhanced through self-adaptation.

The hardware is by definition heterogeneous and hence complex but must be decoupled from the application. The Programming Model will provide a graphical interface and enable the development, analysis and profiling of the application in order to execute in a low-power environment. The gap between the application programmer and the heterogeneous parallel architecture will be covered by a cross-layer approach presented in the architecture section above. During development time, the cross-layer approach will make use of software patterns to automatically generate the required annotations as well as the runtime libraries/configurations for the target architecture.

Subsequently at runtime, resource and job management where jobs will be assigned to execute on specific hardware resources is performed by the Application Life-Cycle Deployment Engine. Together with the Application Monitoring, the Self-Adaptation Manager, the Programming Model runtime and the underlying optimisation engine, they will provide the necessary functionality to adapt the execution of an application or a set of applications toward a more optimal profile in relation to the targeted trade-offs on energy behaviour and time performance. Pragmatically, the implementation plan proposes to adapt SLURM as well as the Programming Model from the StarS family. SLURM is one of the first middleware for performing resource and job management to provide functionalities that enable power monitoring per node along with energy accounting and power profiling per job which is a very useful feature for the project's requirements. Another important aspect is the support of the non-functional requirements of the application (e.g. security, performance, data locality) while running on different architectures. This reduces the need to make major changes to the application implementation (source code) for different architectures. The Programming Model from the StarS family such as OMPSs or COMPSs work at a higher level of abstraction than SLURM as they target application developers. An important implementation decision will therefore be to determine if the Programming Model and SLURM will integrate or if they will provide complementary development approaches. In both cases, the other architecture component will also require integration effort. For instance, the implementation of the Application Life-Cycle Deployment Manager, the Self-Adaptation Manager, the selected programming models from the StarS family as well as the implementation of the underlying optimisation engine (OscaR \cite{OSCAR}) will need to interface with SLURM to augment it with their new capabilities.

\section{Results and Evaluation}
This research is expected to deliver a number of outcomes:

\begin{itemize}
\item A toolbox based implementation of the reference architecture
\item	Reference software development models and methodologies for best practice
\item	A collection of reusable IDE plugins, programming models and runtimes
\item	An adaptive quality model for holistic system performance
\item	Hardware and software energy models.
\end{itemize}

To validate this research, a heterogeneous parallel architectures testbed is provisioned to host the toolbox and the implementation of two Use Cases.

{\it IoT}: the objective is to demonstrate the capabilities of dynamic reconfiguration of FPGA to allow versatile deployment of algorithms applicable in various IoT business use case such as monitor elderly people at home or TV-broadcasting related applications. The reconfigurable power optimized connected platform aims to develop a versatile platform dedicated to network management and data processing. The platform electronics is built around a reconfigurable FPGA (e.g. K7 of Xilinx) with integrated multicore CPU externally managed by either a small low power processor (e.g. ARM M0) or higher end processor according to the business demand. Based on the type of data to transmit and the QoS associated, the low power processor will reconfigure on the fly the FPGA.

{\it HPC}: the objective is to validate the toolbox for HPC workloads with miniapps gathered from different scenarios that can make use of heterogeneous resources to leverage the trade-offs between performance, power and energy consumption. A miniapp is a condensed partial implementation of the HPC application of interest (e.g. weather forecast) that highlights one or multiple performance aspects that can affect the parent application's codebase. A work-load of miniapps composed of MiniFE, which is memory sensitive and phdMesh, which is compute bound might be more efficiently executed upon one specific heterogeneous platform and would be optimally executed through the self-adaptation provided by the architecture. 

Regarding program transformation, both use cases  start from existing applications made available by industry. It is clear that some code transformation and code re-writing will be required in order to better exploit the power offered by the architecture frame-work. Simple scenarios will only require programmers to annotate their program to associate application tasks and their dependencies to follow the requirements of the programming model. In more complex cases, programmers will initially profile an application to identify the algorithmic portion worth adapting to capture inherent algorithm parallelism. For instance, transforming current C/C++ in OpenCL \cite{OpenCL}, current Java code to interface with APARAPI \cite{APARAPI} may be targeted at the lower level with the aim to compile parts of an application for various type of heterogeneous hardware without the need to write hardware specific code. In particular, The research will rely on existing programming framework technologies such as OpenCL SDKs or ROCCC \cite{ROCCC} to generate hardware specific code as proposed in \cite{Segal:2014}.

\section{Related Work}

Several architectures to support low power computing in heterogeneous environments have been proposed in the literature, including those in research projects such as ALMA \cite{ALMA}, 2PARMA \cite{Silvano:2012}, PEPPHER \cite{PEPPHER} \cite{Bruckschloegl:2014}, EXCESS \cite{EXCESS}, P-SOCRATES \cite{Pinho:2014}, FiPS \cite{FiPS}, HARPA \cite{HARPA} and ADEPT \cite{ADEPT}.

The PEPPHER \cite{PEPPHER} \cite{Bruckschloegl:2014} architecture provides a programming framework for C++ applications that targets heterogeneous many-core processors with the aim of ensuring performance and portability. This is achieved by utilising implementations variants for different types of hardware, which are then selected at runtime. Variants can themselves be parallelized in the most suitable framework. The creation of variants was partly automated by libraries created by “expert” programmers with the use of transformation and compilation techniques. The 2PARMA \cite{Silvano:2012} adopted a different approach by using bytecode with a final stage of optimisation to target more specific architectures. It also includes a runtime management component that adapts the code while it is running. An outcome of this project was the Barbeque Open Source Project \cite{Barbeque}.

It is common for these projects to target specific hardware environments that are deemed to be heterogeneous. ALMA \cite{ALMA} for example utilises annotated Scilab code (a Matlab like language), to target two very specific architectures, namely Recode and KIT Kahrisma. The FiPS project \cite{FiPS} has the goal of reducing the power performance ratio within data-centres by integrating FPGAs and other accelerators in high-performance and low-power heterogeneous computing servers, with a focus on the RECS server architecture.

Aspects of power and time criticality have also featured in existing projects. The P-SOCRATES \cite{Pinho:2014} examines time-criticality and parallelization challenges for executing workload-intensive applications with real-time requirements on top of commercial-off-the-shelf (COTS) platforms based on many-core accelerated architectures. ADEPT \cite{ADEPT} which is another ongoing EU Project aiming to develop a tool that will guide software developers and help them to model and predict the power consumption and performance of parallel software and hardware. Also, time criticality and energy are important in many fields of computing, such as Embedded Systems (ES) and HPC. Several past projects have attempted to bridge the perceived gap between these two fields. The EXCESS project \cite{EXCESS} in this regards looked at programming methodologies to drastically simplify the development of energy-aware applications over a range of computing systems whilst considering performance. The HARPA project \cite{HARPA} and its architecture is another example of this with the overall aim of providing efficient mechanisms to offer performance guarantees in the presence of unreliable heterogeneous systems. Providing both proactive and reactive adaptive mechanisms targeting both embedded and HPC-based systems.

Euroserver \cite{Euroserver} aims to find power-efficient solutions for the future datacentre. It is addressing these challenges in a holistic manner: from the architecture point of view, investigating the use of state-of-the-art low-power ARM processors, taking into account the memory and I/O, all managed by new systems software providing transparent system-wide virtualisation and efficient resource use by cloud applications. 

The reference architecture in this paper goes beyond the current state of the art by tackling self-adaptation of both heterogeneous parallel devices and the applications that make use of them using a wider range of optimization criteria (energy consumption, cost, time criticality). Furthermore, the proposed architecture will provide broader support for a wide range of heterogeneous parallel device resources from small (embedded) to large (HPC datacentre environments) and with varying architectures (SoC, CPU, GPGPU, FPGA, etc.). This support will not only be limited to the runtime environment but also filter up the stack to enable device agnostic deployment and provide capabilities to an application developer, through a range of fully integrated software engineering tools (design time modelling, profiling etc.), which are energy aware.

\section{Conclusion}

This paper has highlighted the importance of providing novel methods and tools to support software developers aiming to optimise energy efficiency resulting from designing, developing, deploying and running software on HPAs while maintaining other quality aspects of software to adequate and agreed levels.

The specification of a proposed architecture has been presented, which includes the architectural roles and scope of the components. This architecture complies with standard HPAs and supports an IDE, an application deployment on HPA environments, and heterogeneous parallel device environments. The design of the various architectural components was described, with emphasis on the requirements in order to support energy efficiency management, which is addressed during the complete lifecycle of an application. Future work includes its implementation and evaluation, which will be showcased considering two industrial application deployment illustrations.

\acks

This work is partly supported by the European Commission under H2020-ICT-20152 contract 687584 - Transparent heterogeneous hardware Architecture deployment for eNergy Gain in Operation (TANGO) project.


\bibliographystyle{abbrvnat}


\end{document}